\documentclass[aps,prl,twocolumn,reprint,amsmath,amssymb,groupedaddress,superscriptaddress]{revtex4-2}

\usepackage{graphicx}
\usepackage{bm}
\usepackage{booktabs}
\usepackage{color}
\usepackage[colorlinks=true]{hyperref}
\usepackage{bookmark}

\begin{document}

\title{Correlation-Driven Orbital-Selective Fermiology and Superconductivity in the Bilayer Nickelate \texorpdfstring{La$_3$Ni$_2$O$_7$}{La3Ni2O7}}
\author{Yong-Yue Zong}
\affiliation{National Laboratory of Solid State Microstructures and Department of Physics, Nanjing University, 210093 Nanjing, China}
\author{Shun-Li Yu}
 \email{slyu@nju.edu.cn}
\affiliation{National Laboratory of Solid State Microstructures and Department of Physics, Nanjing University, 210093 Nanjing, China}
\affiliation{Collaborative Innovation Center of Advanced Microstructures, Nanjing University, Nanjing 210093, China}

\author{Jian-Xin Li}
 \email{jxli@nju.edu.cn}
\affiliation{National Laboratory of Solid State Microstructures and Department of Physics, Nanjing University, 210093 Nanjing, China}
\affiliation{Collaborative Innovation Center of Advanced Microstructures, Nanjing University, Nanjing 210093, China}
\affiliation{Jiangsu Key Laboratory of Quantum Information Science and Technology, Nanjing University, Suzhou 215163, China}

\begin{abstract}
Recent angle-resolved photoemission measurements on La$_3$Ni$_2$O$_7$ have challenged the density-functional-theory-based picture of three Fermi surfaces by revealing that the $d_{z^2}$-derived $\gamma$ band can reside below the Fermi level. Motivated by this discrepancy, we investigate a realistic bilayer two-orbital Hubbard model using time-dependent variational principle (TDVP)-based cluster perturbation theory (CPT), alongside large-scale density matrix renormalization group (DMRG) calculations. Our TDVP-CPT calculations, performed on clusters of up to 16 physical sites, reveal that electronic correlations drive a pronounced orbital-selective reconstruction of the low-energy spectrum: the $d_{z^2}$ spectral weight is progressively depleted, the $\gamma$ band sinks below the Fermi level, and pseudogaps open on the remaining $\alpha$ and $\beta$ bands, leaving Fermi arcs dominated by the $d_{x^2-y^2}$ orbital at strong coupling. Furthermore, large-scale DMRG calculations demonstrate that the leading superconducting correlations evolve consistently with this Fermi surface reconstruction, transitioning from $d_{z^2}$-dominated to $d_{x^2-y^2}$-dominated interlayer spin-singlet pairing while retaining an $s_{\pm}$ structure. Consequently, our results indicate that the disappearance of the $\gamma$ pocket is not detrimental to superconductivity; rather, it signals a correlation-driven shift of the pairing channel mediated by interlayer antiferromagnetism, Hund's coupling, and inter-orbital hybridization.
\end{abstract}
\maketitle


\textit{Introduction.---}The discovery of superconductivity (SC) in bilayer nickelate La$_3$Ni$_2$O$_7$ under pressure, with $T_c$ approaching 80~K~\cite{sunSignaturesSuperconductivity2023, zhangHightemperatureSuperconductivity2024, wangPressureInducedSuperconductivity2024}, has established a new frontier in the search for unconventional high-temperature superconductors.
Density functional theory (DFT) calculations indicate that the low-energy electronic structure is governed by two $e_g$ orbitals ($d_{x^2-y^2}$ and $d_{z^2}$), whose bonding and antibonding combinations produce three bands---$\alpha$, $\beta$, and $\gamma$---crossing the Fermi level~\cite{sunSignaturesSuperconductivity2023, luoBilayerTwoOrbital2023, christianssonCorrelatedElectronic2023, zhangElectronicStructure2023, wangElectronicMagnetic2024}.
The three-Fermi-surface picture has served as the foundation for extensive theoretical efforts addressing the pairing symmetry~\cite{yangPossibleWave2023, liuWavePairing2023, tianCorrelationEffects2024, zhangStructuralPhase2024, zhengWaveSuperconductivity2025, guEffectiveModel2025, xiTransitionWave2025, xiaSensitiveDependence2025, fanSuperconductivityNickelate2024, xuIncommensurateSpinfluctuations2025, jiangHighTemperatureSuperconductivity2024} and pairing mechanism~\cite{yangPossibleWave2023, liuWavePairing2023, qinHighSuperconductivity2023, shenShenEffectiveBiLayer2023, yangInterlayerValence2023, luoHighTCSuperconductivity2024, kanekoPairCorrelations2024, kakoiPairCorrelations2024, xiTransitionWave2025, zhengWaveSuperconductivity2025, shenNumericalStudy2025, fanSuperconductivityNickelate2024, ohTypeIIModel2023, luInterplayTwo2024, luInterlayerCouplingDrivenHighTemperature2024, quBilayerModel2024, chenOrbitalSelective2024, quHundsRule2025} in both weak and strong coupling regimes.
Notably, within weak-coupling approaches, the $d_{z^2}$-derived $\gamma$ pocket has been widely invoked as the key ingredient driving $s_\pm$-wave pairing through inter-Fermi-surface scattering~\cite{yangPossibleWave2023, liuWavePairing2023, zhangStructuralPhase2024, guEffectiveModel2025}.
However, since the high-pressure condition severely limits spectroscopic access, the fermiology of pressurized bulk La$_3$Ni$_2$O$_7$ has remained experimentally elusive.

Recent angle-resolved photoemission spectroscopy (ARPES) experiments on La$_3$Ni$_2$O$_7$ thin films have brought this issue into sharp focus~\cite{shenNodelessSuperconducting2025, wangElectronicStructure2025,yueCorrelatedElectronic2025, sunObservationSuperconductivity2025, ryeeSuperconductivityGoverned2025}.
Two independent measurements from different researchers revealed that the $d_{z^2}$-derived $\gamma$ band lies approximately 70~meV below the Fermi level, with no $\gamma$ pocket detected~\cite{sunObservationSuperconductivity2025, wangElectronicStructure2025}.
While other experiments have reported the $\gamma$ band crossing the Fermi level in different samples~\cite{shenNodelessSuperconducting2025, yueCorrelatedElectronic2025}---potentially due to variations in thickness, composition, strain, interfacial reconstruction, or growth conditions---this discrepancy underscores the importance of studying the presence or absence of the $\gamma$ pocket for understanding the pairing mechanisms and pairing symmetry in La$_3$Ni$_2$O$_7$. Furthermore, the electronic bands observed in ARPES exhibit significant renormalization compared with DFT calculations~\cite{sunObservationSuperconductivity2025, wangElectronicStructure2025}. Therefore, studying the effects of electronic correlations on band renormalization, particularly on the $\gamma$ Fermi surface, and their relationship with superconducting pairing symmetry is crucial for clarifying the pairing mechanism in La$_3$Ni$_2$O$_7$~\cite{fanSuperconductivityNickelate2024, ryeeQuenchedPair2024,ryeeSuperconductivityGoverned2025, stepanovCooperatingMultiorbital2026, liaoOrbitalselectiveElectron2024, chenSuperconductivityBilayer2025}.

In this paper, we systematically investigate the bilayer two-orbital Hubbard model for La$_3$Ni$_2$O$_7$, spanning from weak to strong coupling regimes. Our study is grounded in a complete multi-orbital Hubbard interaction with realistic tight-binding parameters, avoiding simplifications or reductions to effective $t$-$J$ models. We employ a combination of time-dependent variational principle (TDVP)-based cluster perturbation theory (CPT) and the density matrix renormalization group (DMRG) method, focusing on the role of electron correlations in fermiology and superconductivity. For the spectral properties, we utilize TDVP-CPT on clusters containing 16 physical sites---larger than in any previous theoretical studies of this system---allowing for a more accurate treatment of short-range correlations within the two-orbital bilayer geometry. To examine the changes in superconducting pairing symmetry alongside band variations, we conduct large-scale DMRG calculations.

Our results reveal a consistent physical picture linking correlation-driven changes in fermiology to the evolution of superconducting pairing symmetry. The TDVP-CPT spectral functions demonstrate pronounced orbital-selective correlation effects. As interactions increase, the $d_{z^2}$-derived $\gamma$ band sinks below the Fermi level while the $\beta$ band lifts, driving successive Lifshitz transitions. At strong coupling, once the heavily renormalized $\gamma$ band is pushed fully below the Fermi level, pseudogaps open on the $\alpha$ and $\beta$ bands, leaving behind Fermi arcs dominated by the $d_{x^2-y^2}$ orbital. Correspondingly, DMRG calculations show that the dominant superconducting correlations undergo a crossover from the $d_{z^2}$ orbital in the weak coupling regime to the $d_{x^2-y^2}$ orbital in the strong coupling regime, while maintaining an $s_{\pm}$-wave pairing symmetry in both cases. In the weak coupling regime, inter-orbital hybridization dominates over Hund's coupling in driving superconductivity by extending the superconducting coherence length. Conversely, in the strong coupling regime, the correlation-driven localization of $d_{z^2}$ electrons establishes a robust interlayer antiferromagnetic exchange background, enabling the itinerant $d_{x^2-y^2}$ electrons to acquire effective pairing interactions via both Hund's coupling and inter-orbital hybridization.

\textit{Model.---}We consider a bilayer model of two $e_g$ orbitals, $d_{x^2-y^2}$ and $d_{z^2}$, 
with its Hamiltonian $H=H_0+H_U$.  The tight-binding term $H_0$ is given by:
\begin{equation}
H_0 \!=\!\!\sum_{ijlm\alpha\beta\sigma }t_{ij,lm}^{\alpha\beta}c^\dagger_{il\alpha\sigma}c_{jm\beta\sigma} + \sum_{il\alpha\sigma}(\epsilon_{\alpha}-\mu) c^\dagger_{il\alpha\sigma}c_{il\alpha\sigma},
	\label{eq:tb}
\end{equation}
where $c^\dagger_{il\alpha\sigma}$ denotes the creation operator for an electron with spin $\sigma$ in orbital $\alpha$ at site $i$ on layer $l$, $\epsilon_\alpha$ is the on-site energy of orbital $\alpha$, and $\mu$ the chemical potential.
The interaction Hamiltonian is $H_U = \sum_i V_i$, where the on-site interaction $V_i$ contains intra-orbital Hubbard term $U$, inter-orbital repulsion $U'$ and Hund's coupling $J_H$:
\begin{align}
	V_i &= \sum_{\alpha} U n_{i\alpha\uparrow} n_{i\alpha\downarrow}
+ \sum_{\alpha < \beta}\sum_{\sigma \sigma'} \left(U' - \frac{J_H}{2} \right) n_{i\alpha\sigma} n_{i\beta\sigma'} \notag
\\
&-\sum_{\alpha < \beta} 2J_H \mathbf{S}_{i\alpha} \cdot \mathbf{S}_{i\beta}
+ \sum_{\alpha \neq \beta} J_H c^{\dagger}_{i\alpha\uparrow} c^{\dagger}_{i\alpha\downarrow} c_{i\beta\downarrow} c_{i\beta\uparrow}.
\label{eq:int}
\end{align}
To simulate  La$_{3}$Ni$_2$O$_7$, we adopt tight-binding parameters from Ref.~\onlinecite{luoBilayerTwoOrbital2023}, and set the average filling at $n=3/8$.
 Regarding the interaction parameters, $U\approx 3.77$ eV is suggested by the constrained random phase approximation ~\cite{christianssonCorrelatedElectronic2023, yueCorrelatedElectronic2025}. While $U= 5.9$ eV is estimated using the linear-response method~\cite{sunSignaturesSuperconductivity2023} which is commonly used in other DFT+$U$ treatments~\cite{liaoOrbitalselectiveElectron2024}.
We use $U=1.5$, 2.2, 3.7 and 6.0 eV to explore the system's behavior across different coupling regimes.

\begin{figure}
\centering
\includegraphics[width=1\linewidth]{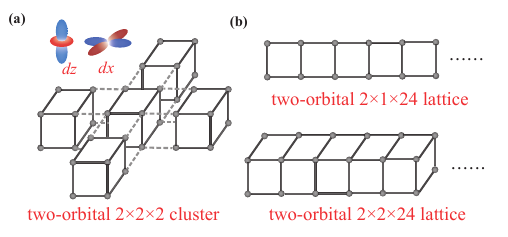}
\caption{ Bilayer two-orbital ($dz$, $dx$ are short for $d_{z^2}$, $d_{x^2-y^2}$ orbitals) lattice structures used in calculations.
(a) It is tiled with small clusters in the TDVP-CPT approach, where DMRG and TDVP are applied to the two-orbital $2\times 2\times 2$  cluster. Solid and dashed lines represent intra- and inter-cluster bonds, respectively.
(b) Large-scale MRG calculations are conducted on the two-orbital $2\times L_y\times L_x$ lattices with $L_y=1, 2$ in studying the superconducting pairing.
}
\label{fig:model}
\end{figure}
\textit{TDVP-CPT.---}In order to study the spectral properties of the system, we employ TDVP-based CPT to solve the model Hamiltonian numerically.
TDVP is a state-of-the-art time-evolution matrix product states (MPS) algorithm\cite{haegemanTimeDependentVariational2011, haegemanUnifyingTime2016, paeckelTimeevolutionMethods2019, vanderstraetenTangentspaceMethods2019}, which has been successfully applied to a wide range of quantum many-body systems\cite{tianMatrixProduct2021, LiTangentSpace2023, shermanSpectralFunction2023, gaoDoubleMagnon2024, wangSpectralProperties2024}.
To calculate the dynamic correlation function:
\begin{equation}
	G_{il\alpha \sigma, jm\beta\sigma}(t) = -i \langle 0|c_{il\alpha \sigma} e^{-i\hat{H}t}c_{jm\beta\sigma}^\dagger|0\rangle e^{iE_0t},
\end{equation}
TDVP is used to evolve the MPS from $c_{jm\beta\sigma}^\dagger|0\rangle$ to  $e^{-i\hat{H}t}c_{jm\beta\sigma}^\dagger|0\rangle$, where $|0\rangle$ and $E_0$ are the ground state and its energy, respectively.
Due to the two-orbital bilayer structure, evolving a finite MPS for a large lattice size is computationally expensive.
Therefore, we perform TDVP on a two-orbital $2\times 2\times 2$ cluster [Fig.~\ref{fig:model}(a)]  to obtain the cluster Green's function $G_{ij}(\omega)=\int_0^t G_{ij}(t) e^{i\omega t}dt$, and approximate the Green's function for the original lattice using CPT\cite{senechalSpectralWeight2000, senechalClusterPerturbationTheory2002, senechalHotSpots2004, yuMottPhysics2011, yuChiralSuperconducting2012, kohnoMottTransition2012, chenAnomalouslyStrong2021, zongPseudogapWith2026}:
\begin{align}
	&\mathcal{G}_{ij}^{-1}(\boldsymbol{k}, \omega) = G_{ij}^{-1}(\omega) - V_{ij}(\boldsymbol{k})\label{eq:cpt}, 
\end{align}
where $V_{ij}(\boldsymbol{k})$ is the  inter-cluster hoppings, and $\mathcal{G}(\boldsymbol{k}, \omega)$ the single-particle Green's function for the original lattice.
The spectral function is given by $A(\boldsymbol{k}, \omega) = -\mathrm{Im}\ G(\boldsymbol{k}, \omega)/\pi$.
In our TDVP-based CPT calculations, to find the ground state of the cluster, we use DMRG and retain the bond dimension $D=4096$ with truncation errors $\varepsilon\lesssim$ $1\times 10^{-6}$.
For the time evolution of the systems with TDVP, we set the time step $\Delta t=0.05$, the maximum time $t_{\mathrm{max}} = 30$ and truncation dimension  $D=2000$.
When performing Fourier transform, a Gaussian type broadening factor $e^{-\eta^2 t^2}$ is multiplied on the integral with $\eta=0.08$.
The DMRG and TDVP calculations in this work are implemented using the MPSKit library~\cite{Van_Damme_MPSKit_2025}, where non-Abelian symmetry U(1)$_{\mathrm{charge}}$ $\times$ SU(2)$_{\mathrm{spin}}$ is implemented with TensorKit library~\cite{juthoTensorKitjl2025}.

\begin{figure}[htbp!]
\centering
\includegraphics[width=1\linewidth]{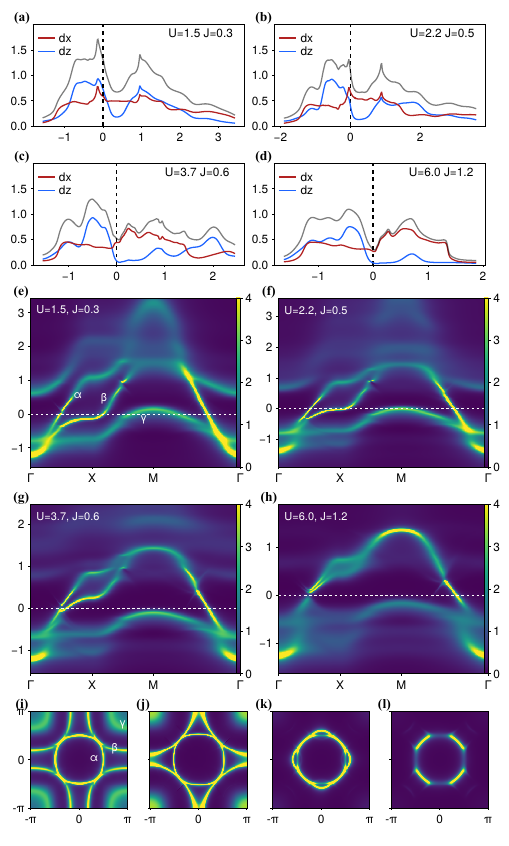}
\caption{
	(a)-(d) show the DOS  calculated with parameters $U=1.5, J_H=0.3$; $U=2.2, J_H=0.5$; $U=3.7, J_H=0.6$ and $U=6.0, J_H=1.2$ (where $U'=U-2J_H$ is used).
	(e)-(h) display the corresponding spectral function along the $\Gamma-X-M-\Gamma$ path in the Brillouin zone.
	The white dashed line is the Fermi level.
	(i)-(l) illustrate the corresponding spectral function in the Brillouin zone at the Fermi level.
}
\label{fig:spectrum}
\end{figure}

\textit{Fermiology.---} Figs.~\ref{fig:spectrum}(a)-(d) illustrate the density of states (DOS) for several typical interaction parameters. It clearly reveals orbital-dependent correlation effects. As the interactions increase, the spectral weights of the $d_{x^2 - y^2}$ and $d_{z^2}$ orbitals at the Fermi level decrease, with the spectral weight of the $d_{z^2}$ orbital diminishing faster. Specifically, the spectral weight of the $d_{z^2}$ orbital at the Fermi level almost disappears at $U = 3.7$ eV [Fig.~\ref{fig:spectrum}(c)], and completely vanishes at $U=6.0$ eV [Fig.~\ref{fig:spectrum}(d)]. In contrast, the $d_{x^2 - y^2}$ orbital retains a finite spectral weight at the Fermi level.

Further insights are provided by the momentum-dependent spectral functions [Figs.~\ref{fig:spectrum}(e)-(l)]. In Fig.~\ref{fig:spectrum}(e), the three bands crossing the Fermi level are denoted as $\alpha, \beta$ and $\gamma$, corresponding to the respective Fermi surface shown in Fig.~\ref{fig:spectrum}(i). In the weak coupling regime ($U=1.5$ eV), the band structure [Fig.~\ref{fig:spectrum}(e)] and Fermi surface [Fig.~\ref{fig:spectrum}(i)] exhibit only slight renormalization relative to the noninteracting system~\cite{luoBilayerTwoOrbital2023}. Upon increasing the interaction strength to $U=2.2$ eV, the $\beta$ band shifts upward, lifting the van Hove singularity near the Fermi level---a feature also corroborated by the noticeable enhancement of the DOS at the Fermi level in Fig.~\ref{fig:spectrum}(b). Crucially, stronger interactions push the $\gamma$ band entirely below the Fermi level [Fig.~\ref{fig:spectrum}(g), $U=3.7$ eV]. This behavior is consistent with recent ARPES measurements on both La$_3$Ni$_2$O$_7$ thin films~\cite{wangElectronicStructure2025, sunObservationSuperconductivity2025} and bulk samples at ambient pressure~\cite{yangOrbitalDependent2024}. Consequently, this interaction-driven band evolution fundamentally alters the Fermi surface topology [Fig.~\ref{fig:spectrum}(k)]. Furthermore, we find that the inter-orbital Coulomb interaction $U^{\prime}$ dominates over Hund's coupling $J_H$ in inducing the upward shift of the $\beta$ band and the elimination of the $\gamma$ pocket [see Fig. S1 in the Supplementary Material (SM)\cite{supplement}].

Notably, as the interaction strength increases, pseudogaps open along the $\Gamma$-$X$ direction, as shown in Figs.~\ref{fig:spectrum}(g) and \ref{fig:spectrum}(h). For moderate interactions ($U=3.7$ eV), the pseudogap initially opens on the $\alpha$ band [Fig.~\ref{fig:spectrum}(g)], leaving only an intact circular Fermi surface [Fig.~\ref{fig:spectrum}(k)]. Under strong interactions ($U=6.0$ eV) [Fig.~\ref{fig:spectrum}(h)], pseudogaps open simultaneously on both the $\alpha$ and $\beta$ bands, resulting in the formation of Fermi arcs [Fig.~\ref{fig:spectrum}(l)]. This pseudogap opening is also clearly visible in the DOS [Figs.~\ref{fig:spectrum}(c) and~\ref{fig:spectrum}(d)]. Furthermore, the orbital-resolved DOS reveals that once the pseudogaps open, the remaining electronic states at the Fermi surface are predominantly of $d_{x^{2}-y^{2}}$ character.

\textit{Superconductivity.---}Next, we employ the DMRG method to explore possible superconducting pairing symmetries as the fermiology evolves. While previous DMRG studies have provided valuable insights based on effective $t$-$J_H$ models or simplified multi-orbital Hubbard models~\cite{quBilayerModel2024, quHundsRule2025, kakoiPairCorrelations2024, chenOrbitalSelective2024, OhType-II2025, shenEffectiveBilayer2023, shenNumericalStudy2025, zhuQuantumPhase2026, jiStrongCoupling2025, chenSuperconductivityBilayer2025}, here we employ the full multi-orbital Hubbard model [see Eqs.~(\ref{eq:tb}) and (\ref{eq:int})], which retains all interaction channels. We perform large-scale DMRG calculations on $2\times L_{y} \times L_{x}$ lattices, with widths $L_{y} \in \{1, 2\}$ and a length of  $L_{x}=24$. We keep a bond dimension up to $D\approx 15000$, yielding truncation errors of $\varepsilon\lesssim1\times 10^{-6}$ for the $L_{y}=1$ system and $\varepsilon\approx1\times 10^{-5}$ for the $L_{y}=2$ system. We define the equal-time pairing correlation function $\Phi(x)$ as follows:
\begin{equation}
	\Phi^{\alpha\beta}_{lm,\hat{\delta}\hat{\delta}'}(x) = \langle \Delta_{lm, \hat{\delta}}^{\alpha\beta}(x_0)^\dagger \Delta_{lm, \hat{\delta}'}^{\alpha\beta}(x_0+x)\rangle,
	\label{eq:pair}
\end{equation}
where $\Delta_{lm, \hat{\delta}}^{\alpha\beta}(x)=[c_{xl\alpha\uparrow} c_{(x+\hat{\delta})m\beta\downarrow}- c_{xl\alpha\downarrow} c_{(x+\hat{\delta})m\beta\uparrow}]/\sqrt{2}$ is the spin-singlet pairing operator. From this, we construct the correlation functions for various pairing symmetries. For instance, the interlayer $s$-wave pairing is given by $\Phi^{\alpha\beta}_{o}=\sum_{l\neq m}\Phi^{\alpha\beta}_{lm,\hat{o}\hat{o}}$ with $\hat{o}=(0,0)$. Similarly, the intralayer $s$-wave and $d$-wave pairings are defined as $\Phi^{\alpha\beta}_{s}=\sum_{l}\Phi^{\alpha\beta}_{ll,\hat{o}\hat{o}}$ and $\Phi^{\alpha\beta}_{d}=\sum_l(\Phi^{\alpha\beta}_{ll,\hat{x}\hat{x}}-\Phi^{\alpha\beta}_{ll,\hat{x}\hat{y}}-\Phi^{\alpha\beta}_{ll,\hat{y}\hat{x}}+\Phi^{\alpha\beta}_{ll,\hat{y}\hat{y}})$, respectively,  where $\hat{x}=(1,0)$ and $\hat{y}=(0,1)$. To minimize boundary effects induced by open boundary conditions, we set the reference point at $x_0={L}/{4}+1$ and restrict the measured coordinates to the bulk region up to ${3L}/{4}$, effectively discarding ${L}/{4}$ sites at each end.

\begin{figure}
\centering
\includegraphics[width=1.\linewidth]{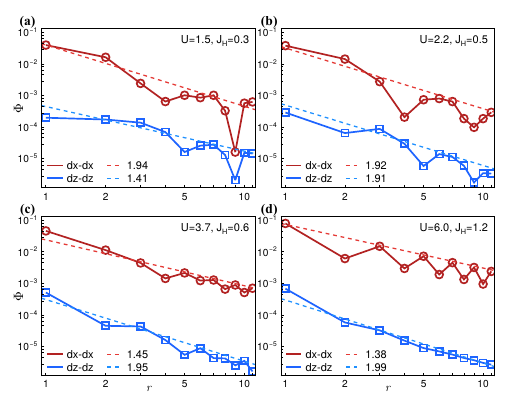}
\caption{
Interlayer pairing correlations $\Phi_{o}^{xx}$ and $\Phi_{o}^{zz}$ calculated on the $2\times 1\times 24$ lattice for (a) $U=1.5, J_H=0.3$; (b) $U=2.2, J_H=0.5$; (c) $U=3.7, J_H=0.6$ and (d) $U=6.0, J_H=1.2$, where $U'=U-2J_H$ is used.
The dashed lines are power-law fits with $\Phi\sim r^{-K_{sc}}$, where the values of $K_{sc}^{(dx-dx)}$ and $K_{sc}^{(dz-dz)}$ are labeled.
}
\label{fig:sc}
\end{figure}
Figure~\ref{fig:sc} presents the interlayer pairing correlations $\Phi_{o}^{xx}$ and $\Phi_{o}^{zz}$ on the $2\times 1\times 24$ lattice for the same interaction parameters used in the preceding spectral analysis, allowing us to directly study the changes in pairing symmetry as the electronic structure evolves. The decay of these correlations exhibits an algebraic behavior $\Phi \sim r^{-K_{sc}}$, where $K_{sc}<2$ implies divergent pairing susceptibility $\chi_{sc}\sim1/T^{(2-K_{sc})}$ at low temperatures, signaling quasi-long-range superconducting order. At weak coupling ($U=1.5$ eV), where the $\gamma$ Fermi surface is still present, we find $K_{sc}^{(dz-dz)}=1.41$ and  $K_{sc}^{(dx-dx)}=1.94$ as shown in Fig.~\ref{fig:sc}(a), indicating that superconductivity is predominantly driven by the $d_{z^2}$ orbital. This is consistent with previous weak-coupling theory predictions that the $\gamma$ Fermi surface favors $d_{z^{2}}$ orbital dominated superconductivity~\cite{yangPossibleWave2023, liuWavePairing2023, yangInterlayerValence2023, xiTransitionWave2025}. As the interaction strength increases and the $\gamma$ band sinks below the Fermi level, superconductivity in the $d_{z^2}$ orbital weakens, with $K_{sc}^{(dz-dz)}$ growing from 1.4(1) at $U=1.5$ eV to 1.99 at $U=6.0$ eV [Fig.~\ref{fig:sc}(d)]. Conversely, the superconducting correlation in the $d_{x^2-y^2}$ orbital becomes progressively stronger, with $K_{sc}^{(dx-dx)}$ decreasing from 1.94 at $U=1.5$ eV to 1.92 at $U=2.2$ eV [Fig.~\ref{fig:sc}(b)], further to 1.45 at $U=3.7$ eV [Fig.~\ref{fig:sc}(c)] and finally to 1.38 at $U=6.0$ eV [Fig.~\ref{fig:sc}(d)]. These results demonstrate that as the $\gamma$ Fermi surface is eliminated by electronic correlations, the system undergoes a transition from $d_{z^{2}}$-dominated to $d_{x^{2}-y^{2}}$-dominated superconductivity.

To establish interlayer pairing as the dominant channel and justify the neglect of intralayer pairings, we investigate a $2\times 2\times 24$ lattice, which serves as a minimal setup with two-dimensional characteristics and accommodates both types of pairing. We find that the intralayer pairings---including the onsite $s$-wave, $d$-wave, and extended $s$-wave---all exhibit exponential decay. In contrast, only the interlayer pairing shows algebraic decay (see Fig. S2 in SM~\cite{supplement}). This confirms that interlayer pairing is indeed the dominant superconducting channel in this model.

Furthermore, the bilayer structure of model~(\ref{eq:tb}) possesses a mirror symmetry relating the top and bottom layers. Consequently, each Fermi surface corresponds to a bonding (symmetric) or antibonding (antisymmetric) combination of electronic wavefunctions from the two layers. Specifically, the $\alpha$ and $\gamma$ pockets are derived from the bonding states, while the $\beta$ pocket is derived from the antibonding states. The interlayer pairing causes the superconducting pairing functions on the bonding and antibonding pockets to have opposite signs. Therefore, based on the sign distribution of the pairing functions across the Fermi surfaces, the interlayer spin-singlet pairing gives rise to $s_{\pm}$-wave superconductivity.

\begin{figure}
\centering
\includegraphics[width=1\linewidth]{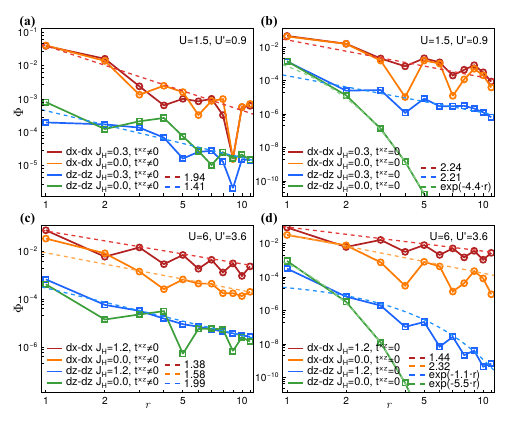}
\caption{
Pairing correlations with fixed $U$ and $U'$ on the $2\times 1\times 24$ lattice.
$\Phi_{o}^{xx}$ and $\Phi_{o}^{zz}$ for (a) $U=1.5,U'=0.9$ with various $J_H$.
(b) $U=1.5,U'=0.9$ with various $J_H$ with $t^{xz}=0$.
(c) $U=6,U'=3.6$ with various $J_H$.
(d) $U=6,U'=3.6$ with various $J_H$ with $t^{xz}=0$.
$t^{xz}\neq 0$  means that $t^{xz}_3=0.239$ and $t^{xz}_4=-0.034$ extracted from Ref.~\cite{luoBilayerTwoOrbital2023}.
}
\label{fig:sc2}
\end{figure}

Figure~\ref{fig:sc2} illustrates how the pairing correlation functions, $\Phi_{o}^{xx}$ and $\Phi_{o}^{zz}$, depend on Hund's coupling ($J_H$) and inter-orbital hybridization ($t^{xz}$) across both weak and strong coupling regimes. As depicted in Fig.~\ref{fig:sc2}(a), within the weak-coupling regime ($U=1.5$ eV), variations in $J_H$ have a negligible impact on the pairing correlation functions. Conversely, a reduction in $t^{xz}$ leads to a substantial suppression of pairing correlations in both the $d_{x^2-y^2}$ and $d_{z^2}$ orbitals. Figure~\ref{fig:sc2}(b) demonstrates that for $t^{xz}=0$, $K_{sc}>2$ occurs in both orbitals; notably, $\Phi_{o}^{zz}$ even exhibits an exponential decay when both $t^{xz}$ and $J_H$ vanish. These findings suggest that inter-orbital hybridization is the dominant factor in driving superconductivity within the weak-coupling regime. Specifically, while superconductivity in this regime is predominantly driven by the $d_{z^2}$ orbital, its minimal in-plane intra-orbital hopping results in a short Cooper-pair coherence length. Consequently, the primary role of inter-orbital hybridization is to extend this coherence length, thereby robustly enhancing superconductivity.

In contrast, within the strong-coupling regime, where superconductivity is governed by the $d_{x^2-y^2}$ orbital, reducing either $J_H$ or $t^{xz}$ individually fails to significantly suppress the superconducting state. As shown in Fig.~\ref{fig:sc2}(c) for $U=6$ eV, decreasing $J_H$ from $1.2$ eV to $0$ only moderately weakens the pairing correlation in the $d_{x^2-y^2}$ orbital, with $K_{sc}$ increasing from $1.38$ to $1.58$. Similarly, fixing $J_H=1.2$ eV while reducing $t^{xz}$ to zero causes a marginal shift in $K_{sc}$ from $1.38$ to $1.44$ [Figs.~\ref{fig:sc2}(c) and~\ref{fig:sc2}(d)]. However, simultaneously eliminating both $J_H$ and $t^{xz}$ results in a rapid decay of $\Phi_{o}^{xx}$, yielding $K_{sc}=2.32$. This demonstrates that both Hund's coupling and inter-orbital hybridization are essential for superconductivity in the strong-coupling regime. Notably, Hund's coupling appears to exert a stronger influence, given that its elimination produces a larger variation in $K_{sc}$. As in the weak-coupling case, the interlayer pairing amplitude of the $d_{z^{2}}$ orbital remains the largest among all pairing channels in the strong-coupling regime (see Tables S1 and S2 in SM~\cite{supplement}). Nevertheless, its pairing correlation decays significantly faster than that of the $d_{x^{2}-y^{2}}$ orbital [Fig.~\ref{fig:sc2}(c)]. This rapid decay arises because the nearly half-filled $d_{z^2}$ electrons become heavily localized in the strong-coupling regime. This is fully consistent with the orbital-selective Mott physics observed in our spectral analysis, which reveals a near absence of $d_{z^2}$ spectral weight around the Fermi level [see Fig.~\ref{fig:spectrum}(d)]. Instead, the itinerant $d_{x^{2}-y^{2}}$ electrons leverage the interlayer antiferromagnetic exchange of the $d_{z^2}$ orbitals to establish superconducting pairing via Hund's coupling and inter-orbital hybridization~\cite{luInterlayerCouplingDrivenHighTemperature2024, quBilayerModel2024,4qsd-ws9c}. This mechanism is further corroborated by the enhancement of the $d_{x^{2}-y^{2}}$ interlayer pairing amplitude upon increasing $J_H$ and $t^{xz}$ (see Table S2 in SM~\cite{supplement}). Furthermore, we observe that the complete suppression of Hund's coupling drives the system toward a charge density wave (see Fig.~S4 in SM~\cite{supplement}), indicating that $J_H$ further stabilizes superconductivity by suppressing this competing instability.

\textit{Conclusion.---}In conclusion, our study establishes electronic correlations as the common origin for both the reconstructed Fermi surface and the evolving superconducting pairing in bilayer La$_3$Ni$_2$O$_7$. Using TDVP-CPT, we identify a strong orbital-selective renormalization. Increasing interactions deplete the coherent low-energy spectral weight of the $d_{z^2}$ electrons and push the $\gamma$ band below the Fermi level. This process opens pseudogaps on the $\alpha$ and $\beta$ bands, yielding $d_{x^2-y^2}$-dominated Fermi arcs in the strong-coupling regime. Furthermore, DMRG calculations on the same multi-orbital Hubbard model demonstrate that this Fermi surface reconstruction is accompanied by a crossover in the pairing channel. The leading superconducting pairing correlation shifts from the $d_{z^2}$ orbital to the $d_{x^2-y^2}$ orbital, preserving an $s_{\pm}$ sign structure governed by the bonding and antibonding bands. The microscopic mechanism of superconductivity is highly regime dependent. At weak coupling, inter-orbital hybridization drives the  $d_{z^2}$ orbital pairing by extending the superconducting coherence length, dominating over Hund's coupling. Conversely, at strong coupling, the correlation-driven localization of $d_{z^2}$ states provides a robust interlayer antiferromagnetic background. This environment enables the itinerant $d_{x^2-y^2}$ electrons to develop effective pairing interactions through the combined effects of Hund's coupling and inter-orbital hybridization.

\begin{acknowledgments}
We gratefully acknowledge discussions with Zhao-Long Gu and Li-Wei He. This work was supported by National Key Projects for Research and Development of China (No. 2024YFA1408104 and No. 2021YFA1400400), National Natural Science Foundation of China (No. 12374137, No. 12434005, and No. 12550405).
We thank e-Science Center of Collaborative Innovation Center of Advanced Microstructures for support in allocation of CPU.
\end{acknowledgments}

\bibliographystyle{apsrev4-2}
\bibliography{refs}

\end{document}


\title{Supplemental Material for ``Correlation-Driven Orbital-Selective Fermiology and Superconductivity in the Bilayer Nickelate \texorpdfstring{La$_3$Ni$_2$O$_7$}{La3Ni2O7}"}

\author{Yong-Yue Zong}
\affiliation{National Laboratory of Solid State Microstructures and Department of Physics, Nanjing University, 210093 Nanjing, China}
\author{Shun-Li Yu}
 \email{slyu@nju.edu.cn}
\affiliation{National Laboratory of Solid State Microstructures and Department of Physics, Nanjing University, 210093 Nanjing, China}
\affiliation{Collaborative Innovation Center of Advanced Microstructures, Nanjing University, Nanjing 210093, China}

\author{Jian-Xin Li}
 \email{jxli@nju.edu.cn}
\affiliation{National Laboratory of Solid State Microstructures and Department of Physics, Nanjing University, 210093 Nanjing, China}
\affiliation{Collaborative Innovation Center of Advanced Microstructures, Nanjing University, Nanjing 210093, China}
\affiliation{Jiangsu Key Laboratory of Quantum Information Science and Technology, Nanjing University, Suzhou 215163, China}

\maketitle

\section{inter-orbital repulsion interaction}
\begin{figure}[!tbp]
	\centering
		\includegraphics[width=1\linewidth]{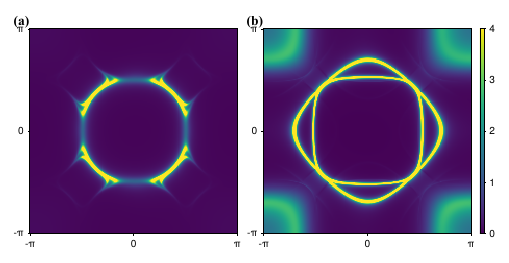}
\caption{ Spectral function in the first Brillouin zone at Fermi level with
(a) $U=6, U'=3.6$ and $J=0$, (b) $U=6, U'=0$ and $J=0$.
}
\label{fig:fs}
\end{figure}

In Fig.~\ref{fig:fs}, we present the spectral functions in the first Brillouin zone at the Fermi level with $U=6$.
To isolate the effects of interaction parameters, we suppress the values of $J$ and $U'$.
In fig.~\ref{fig:fs}(a), with $J=0$, the Fermi surfaces show slight changes compared to those when $J=1.2$ in the main text.
The downward shift of the $\gamma$ band and the emergence of the pseudogap remain evident. When both $J=0$ and $U'=0$, the $\gamma$ pocket is maintained, and the pseudogap is absent. Thus, we conclude that the inter-orbital repulsion $U'$ significantly contributes to the downward shift of the $\gamma$ band and the formation of the pseudogap. Additionally, we observe that the change in the $\beta$ pocket topology has already occurred, suggesting that the upward shift of the $\beta$ pocket might result from the intra-orbital repulsion $U$.

\begin{figure}[!tbp]
	\centering
		\includegraphics[width=0.96\linewidth]{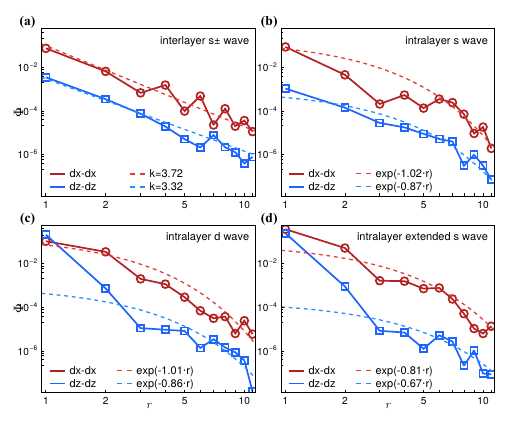}
\caption{Pairing correlation functions for 
(a) interlayer $s\pm$ wave symmetry, (b) intralayer $s$ wave symmetry, (c) intralayer $s$ wave symmetry and (d) intralayer extended-$s$ wave symmetry on the $2\times 2\times 24$  lattice with $U=1.5, J_H=0.3$.
}
\label{fig:sc}
\end{figure}

\begin{figure}[!tbp]
	\centering
		\includegraphics[width=0.96\linewidth]{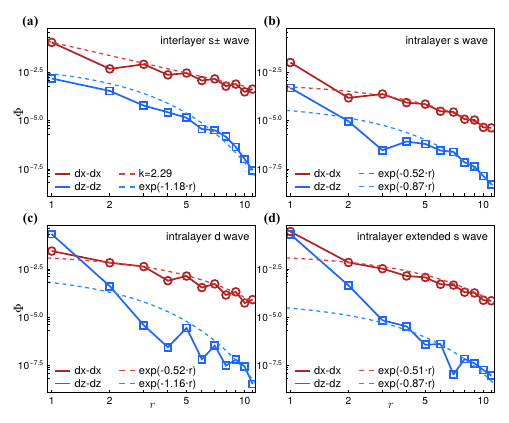}
\caption{Pairing correlation functions in
(a) interlayer $s\pm$ wave symmetry, (b) intralayer $s$ wave symmetry, (c) intralayer $s$ wave symmetry and (d) intralayer extended-$s$ wave symmetry on the $2\times 2\times 24$  lattice with $U=6.0, J_H=1.2$.
}
\label{fig:sc2}
\end{figure}

\begin{table}[!tbp]
\centering
\caption{Pairing correlation functions with $U=1.5$, $U'=0.9$ at $r=0$ on $2\times 1\times 24$ lattice}
\begingroup
\setlength{\tabcolsep}{4pt}
\renewcommand{\arraystretch}{0.92}
\begin{tabular}{lcc}
\toprule
Condition & $\Phi_o^{xx}(0)$ & $\Phi_o^{zz}(0)$ \\
\midrule
$t^{xz}\neq 0$, $J_H=0.3$ & 0.0720 & 0.6486 \\
$t^{xz}\neq 0$, $J_H=0.0$ & 0.0649 & 0.6287 \\
$t^{xz}=0$,   $J_H=0.3$   & 0.0735 & 0.7255 \\
$t^{xz}=0$,   $J_H=0.0$   & 0.0650 & 0.7275 \\
\bottomrule
\end{tabular}
\endgroup
\end{table}

\begin{table}[!tbp]
\centering
\caption{Pairing correlation functions with $U=6.0$, $U'=3.6$ at $r=0$ on $2\times 1\times 24$ lattice}
\begingroup
\setlength{\tabcolsep}{4pt}
\renewcommand{\arraystretch}{0.92}
\begin{tabular}{lcc}
\toprule
Condition & $\Phi_o^{xx}(0)$ & $\Phi_o^{zz}(0)$ \\
\midrule
$t^{xz}\neq 0$, $J_H=1.2$ & 0.1888 & 0.7138 \\
$t^{xz}\neq 0$, $J_H=0.0$ & 0.0721 & 0.8593 \\
$t^{xz}=0$,   $J_H=1.2$   & 0.1643 & 0.7797 \\
$t^{xz}=0$,   $J_H=0.0$   & 0.0605 & 0.9353 \\
\bottomrule
\end{tabular}
\endgroup
\end{table}

\section{Pairing correlation functions}
In Fig.~\ref{fig:sc} and Fig.~\ref{fig:sc2}, we present the pairing correlation functions of $U=1.5, J_H=0.3$ and $U=6.0, J_H=1.2$ on the $2\times 2\times 24$  lattice using density matrix renormalization group (DMRG). We observe that only the interlayer pairing correlation function exhibits algebraic decay as shown in Fig.~\ref{fig:sc}(a) and Fig.~\ref{fig:sc2}(a), while the onsite $s$-wave, $d$-wave, and extended $s$-wave---all exhibit exponential decay, which demonstrates that the interlayer pairing is clearly the dominant channel compared with the other channels in Fig.~\ref{fig:sc}(b-d) and Fig.~\ref{fig:sc2}(b-d). We note that the extracted Luttinger exponent satisfies $K_{\mathrm{sc}} > 2$. Within the standard DMRG/Luttinger-liquid criterion, quasi-long-range superconducting order would require $K_{\mathrm{sc}} < 2$; therefore, the present $L_y = 2$ data do not constitute evidence for quasi-long-range superconducting order in that channel, but rather indicate only short-range pairing correlations. We attribute this discrepancy to limited numerical accuracy imposed by computational constraints: for the $2\times 2\times24$  system we can only keep a relatively modest bond dimension ($D \approx 15000$), which is likely insufficient to fully capture the long-distance behavior. In contrast, for the $2\times 1\times24$ system presented in the main text, the same pairing channel shows behavior consistent with quasi-long-range order and thus supports the expectation that, with substantially increased computational resources (larger $D$) for $L_y = 2$, the dominant pairing channel may ultimately develop quasi-long-range superconducting correlations as well.

\begin{figure}[!htbp]
	\centering
		\includegraphics[width=0.96\linewidth]{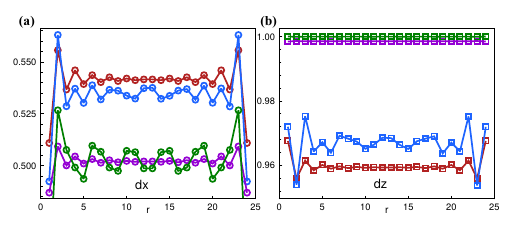}
\caption{Charge density distributions with $U=6$ in
(a) $d_{x^2-y^2}$ orbital, (b) $d_{z^2}$ orbital.
The different color of the curves represent different model parameters: the red curves correspond to $J_H=1.2, t^{xz}\neq 0$; the blue curves correspond to $J_H=0.0, t^{xz}\neq 0$; the purple curves correspond to $J_H=1.2, t^{xz}= 0$ and the green curves correspond to $J_H=0.0, t^{xz}= 0$.
}
\label{fig:cdw}
\end{figure}
\section{Charge density distribution}
In Fig.~\ref{fig:cdw}, we present the charge density distributions obtained from density matrix renormalization group (DMRG) calculations, using model parameters consistent with those in Fig.4(c) and Fig.4(d) of the main text. We observe that the charge density wave (CDW) intensifies as $J_H$ is reduced. Furthermore, when both $J_H$ and $t^{xz}$ are simultaneously suppressed, the CDW reaches its maximum strength among the systems.